\documentclass{article}
\usepackage{iclr2027_conference,times}

\usepackage{hyperref}
\usepackage{url}
\hypersetup{
  colorlinks=false,
  pdfborder={0 0 0.45},
  linkbordercolor={0.35 0.35 0.35},
  citebordercolor={0.35 0.35 0.35},
  urlbordercolor={0.35 0.35 0.35},
  pdftitle={When Do Tensor Product Paths Matter? A Group-Theoretic Analysis of Equivariant Interatomic Potentials},
  pdfauthor={Jiawei Wang; Sai Gu; Laurence T. Wang; Junfu Tan; Yemin Yu; Ziheng Lu},
  pdfsubject={Preprint}
}
\usepackage{amssymb}
\usepackage{amsthm}
\usepackage{booktabs}
\usepackage{multirow}
\usepackage{graphicx}
\usepackage{float}
\usepackage{xcolor}
\usepackage{microtype}
\usepackage[ruled,vlined,linesnumbered]{algorithm2e}

\usepackage{amsmath,amsfonts,bm}

\def\eqref#1{equation~\ref{#1}}

\def\1{\bm{1}}

\DeclareMathAlphabet{\mathsfit}{\encodingdefault}{\sfdefault}{m}{sl}
\SetMathAlphabet{\mathsfit}{bold}{\encodingdefault}{\sfdefault}{bx}{n}

\newcommand{\DMAE}{\Delta\mathrm{MAE}}      
\newtheorem{proposition}{Proposition}

\title{When Do Tensor Product Paths Matter?\\
       A Group-Theoretic Analysis of\\
       Equivariant Interatomic Potentials}

\author{
  Jiawei Wang\textsuperscript{1,2}\quad
  Sai Gu\textsuperscript{2}\quad
  Laurence T. Yang\textsuperscript{3}\quad
  Junfu Tan\textsuperscript{4}\quad
  Yemin Yu\textsuperscript{1}\quad
  Ziheng Lu\textsuperscript{1}\thanks{Corresponding author} \\
  \textsuperscript{1}Kairos Materials \\
  \textsuperscript{2}School of Engineering, University of Warwick \\
  \textsuperscript{3}Zhengzhou University \\
  \textsuperscript{4}The University of Hong Kong \\
}

\iclrfinalcopy
\begin{document}
\maketitle
\fancyhead[L]{Preprint}

\begin{abstract}
Tensor-product (TP) layers in equivariant interatomic potentials encode angular interactions through
symmetry-allowed paths.  Yet local site symmetry can make an allowed path carry no signal after
neighbor aggregation.  We call this effect \emph{Aggregation Cancellation}.  In a scalar-input
first layer, equal-species neighbors on a complete site-symmetry orbit share scalar and radial
factors, reducing their angular contribution to a spherical-harmonic orbit sum.  If the restricted
degree-$l$ representation has no trivial component, this sum and the corresponding path vanish for
any learned radial weight.  For an ideal $O_h$ octahedral orbit, this removes the
$l_\mathrm{edge}=1,2,3$ classes.  Beyond the first layer, the factorisation no longer holds, so
cancellation is no longer guaranteed; we instead ask whether affected paths remain less important
on average.  We test the two levels separately in NequIP: matched-radius TiO$_6$ probes examine
first-layer cancellation, whereas a within-model CaTiO$_3$ comparison and six OMat24 models examine
the importance bias.  We then rank paths on training frames, choose the pruning fraction on
validation frames, and evaluate frozen masks on held-out tests.  At the highest evaluated fraction
(60\% of 174 paths), force mean absolute error (MAE)
changes are 1.42\%, 0.04\%, and 1.83\% for CaTiO$_3$, OMat24 $O_h$-cubic, and OMat24 low-symmetry
data, respectively.  The masks imply 67.12--71.94\% potential TP-operation reduction and 59.77\%
potential edge multilayer perceptron (MLP) reduction; these are structural proxies, not measured
speedups.  The results
connect local symmetry to path redundancy and pruning robustness in NequIP; transfer to other
architectures remains open.

\end{abstract}

\section{Introduction}
\label{sec:intro}

Equivariant interatomic potentials encode atomic geometry by coupling learned features with
spherical harmonics through Clebsch--Gordan tensor products (TPs)
\citep{batzner2022nequip,batatia2022mace,musaelian2023allegro,liao2023equiformer,
geiger2022e3nn}.  A TP instruction combines an input feature of angular momentum
$l_\mathrm{in}$ with an edge harmonic of degree $l_\mathrm{edge}$ to produce an output feature
$l_\mathrm{out}$.  Triangle and parity rules determine which instructions are allowed by
equivariance, but they do not determine whether neighbor aggregation supplies a nonzero signal to
each allowed instruction.  This distinction is consequential because practical models may contain
hundreds of weighted paths: the five-layer NequIP architecture studied here has 174 at
$l_{\max}=3$.

The unresolved question is therefore not only which TP paths an architecture permits, but when
those paths matter for the environments presented to a trained model.  Recent methods reduce TP
cost by changing the computation, selecting paths architecturally, or learning gates
\citep{passaro2023reducing,xie2025price,lee2025flashtp,luo2025sphnet}; pruning criteria instead
rank parameters by magnitude or sensitivity \citep{lecun1990obd,molchanov2019importance}.  These
approaches can expose redundancy, but a low score does not by itself explain why a path is inactive
on a particular neighbor distribution.  Likewise, a favorable curve evaluated on the data used to
choose a mask does not establish that the mask generalises.  A mechanistic account therefore needs
both a symmetry-based prediction and an evaluation that separates ranking, selection, and test.

We address this gap by analysing neighbor aggregation under local site symmetry.  In a scalar-input
first layer, same-species neighbors begin with the same embedding.  Members of one site-symmetry
orbit also share a radius and therefore a radial-network output.  Their angular contribution
factorises into the spherical-harmonic orbit sum
$\mathbf S_l=\sum_jY_l(\hat{\mathbf r}_{ij})$.  Because the site group only permutes the summands,
$\mathbf S_l$ lies in the trivial subspace of the restricted degree-$l$ representation.  If this
subspace is absent, the orbit sum and the associated first-layer TP path vanish for any learned
radial weight.  We call this effect \emph{Aggregation Cancellation}.

This argument produces an exact first-layer result and a testable hypothesis for the full network.
For an ideal $O_h$ octahedral orbit, the $l_\mathrm{edge}=1,2,3$ first-layer contributions cancel
exactly.  In deeper layers, symmetry-related neighbors carry transformed equivariant features
rather than a shared scalar, so the factorisation no longer holds in general.  The exact selection
rule therefore does not imply permanent inactivity.  Instead, we test whether the first-layer
bottleneck appears as a statistical bias toward lower path importance.  Conversely, lowering site
symmetry permits more invariant components but does not rank their learned magnitudes; bonding,
composition, distortion, and temperature remain relevant.

We test the exact and statistical claims separately.  Matched-radius TiO$_6$ probes test
first-layer cancellation and parity-dependent activation under controlled distortions.  A
within-model CaTiO$_3$ temperature comparison and six OMat24 NequIP models test the broader
importance bias and its limits.  Finally, a three-system pruning study ranks paths on training
frames, chooses the pruning fraction on validation frames, and evaluates the frozen mask on
held-out test frames; post hoc controls are kept separate.  All trained-model evidence uses
NequIP.  The first-layer argument may apply to other architectures that satisfy the same scalar-input
and radial-aggregation assumptions, but their importance and pruning behaviour are not tested here.

\paragraph{Contributions.}
\begin{itemize}
  \item We derive an orbit-factorisation result and a group-independent criterion, expressed using
        the trivial irrep $\mathbf1_{G_i}$, for exact first-layer TP-path cancellation under
        complete equal-species, equal-radius neighbor orbits.
  \item Controlled TiO$_6$ probes test the exact first-layer result, while a within-model
        CaTiO$_3$ comparison and six OMat24 NequIP models test the predicted importance bias without
        claiming that symmetry alone orders path magnitude.
  \item Under disjoint ranking, selection, and test roles, the validation-selected 60\% masks change
        held-out force mean absolute error (MAE) by 0.04--1.83\% across three systems.  They imply
        67.12--71.94\% potential TP-operation reduction and 59.77\% potential edge multilayer
        perceptron (MLP) reduction; these are structural operation counts, not measured speedups.
        Post hoc OMat24 controls compare the gradient masks with equal-count random and magnitude
        baselines.
\end{itemize}

\section{Related Work}
\label{sec:related}

\paragraph{Equivariant interatomic potentials.}
Tensor field networks and related constructions introduced learned features that transform under
irreducible representations of rotations \citep{cohen2016group,kondor2018generalization,
thomas2018tensor,anderson2019cormorant}.  The e3nn library made these representation-theoretic
operations reusable across architectures \citep{geiger2022e3nn}, while SO3krates developed an
alternative equivariant attention-based atomistic representation \citep{frank2022so3krates}.  NequIP
showed that such features can produce data-efficient force fields, while MACE connected equivariant
message passing to higher-body-order atomic cluster expansions \citep{batzner2022nequip,
batatia2022mace}.  Allegro localised the representation for scalable simulation, and EquiformerV2
combined equivariant operations with attention \citep{musaelian2023allegro,liao2023equiformer}.
SevenNet further demonstrates scalable equivariant potentials for periodic materials
\citep{park2024sevennet}.
Unified analyses relate these models to the atomic cluster expansion and clarify their expressivity
\citep{drautz2019ace,nigam2022unified,pozdnyakov2022incompleteness}.  These works establish what an
architecture can represent; our question is when the neighbor distribution annihilates a specific
allowed path before its learned coefficient can matter.

\paragraph{Efficient tensor products.}
Several recent methods reduce TP cost by changing the computation itself.  SO(3)-to-SO(2)
reductions exploit rotations into an edge-aligned frame \citep{passaro2023reducing}; path-design
studies remove couplings according to architectural criteria \citep{xie2025price}; and fused or
low-rank implementations improve kernel efficiency \citep{lee2025flashtp}.  Adaptive path gates
have also been learned for equivariant Hamiltonian prediction \citep{luo2025sphnet}.  These methods
are complementary to our analysis.  Aggregation Cancellation predicts that the utility of a path
depends on the site-symmetry orbits contained in the deployment data.  It can therefore guide
structural path removal, but only after its distributional assumptions are checked.

\paragraph{Symmetry adaptation and pruning.}
Point-group-adapted representations explicitly restrict features to subspaces compatible with a
crystal symmetry \citep{heilman2026pointgroup}.  We instead show that an ordinary $E(3)$-equivariant
message-passing layer can encounter the same trivial-subspace projection implicitly through orbit
aggregation.  Classical pruning scores use weight magnitude, second-order sensitivity, or gradient
criteria \citep{lecun1990obd,molchanov2019importance,frankle2019lottery}.  Our gradient score is a
practical ranking device rather than the theoretical contribution.  The group argument explains
why the data may produce low sensitivity, while the disjoint ranking/selection/test protocol tests
whether that sensitivity generalises.

\section{Aggregation Cancellation}
\label{sec:theory}

\subsection{First-Layer Orbit Factorisation}

Consider a center atom $i$ and a weighted TP instruction
$p=(l_\mathrm{in},l_\mathrm{edge},l_\mathrm{out})$.  Suppressing channel indices, its message is
\begin{equation}
M^{l_\mathrm{out}}_{i,m}
=\sum_{j\in\mathcal N(i)}w_p(r_{ij})
\sum_{m_\mathrm{in},m_\mathrm{edge}}
C^{m}_{m_\mathrm{in},m_\mathrm{edge}}
h^{l_\mathrm{in}}_{j,m_\mathrm{in}}
Y^{m_\mathrm{edge}}_{l_\mathrm{edge}}(\hat{\mathbf r}_{ij}).
\label{eq:message_main}
\end{equation}
Here $w_p$ is produced by the edge MLP, $C$ is a Clebsch--Gordan coefficient
\citep{wigner1959theory,varshalovich1988quantum}, and
$Y_l^m$ is a spherical harmonic.  Let $G_i$ be the site-symmetry group of the center and partition
a complete neighbor shell into orbits $\mathcal O_a$ under $G_i$.  Within one orbit, symmetry
preserves radius and species.

\begin{proposition}[Orbit factorisation]
For a first-layer path with $l_\mathrm{in}=0$, the contribution of orbit $\mathcal O_a$ factorises as
\begin{equation}
M^{l_\mathrm{out}}_{i,m}(\mathcal O_a)
=\alpha_{a,p}\,S^{(a)}_{l_\mathrm{edge},m},\qquad
S^{(a)}_{l,m}=\sum_{j\in\mathcal O_a}Y_l^m(\hat{\mathbf r}_{ij}),
\label{eq:factor_main}
\end{equation}
where $\alpha_{a,p}$ contains the common species feature, radial weight, channel mixing, and CG
factor.  Hence $\mathbf S_l^{(a)}=\mathbf0$ makes the whole orbit contribution zero for any learned
radial weight.
\end{proposition}

The assumptions are explicit.  First, the input is the scalar species embedding used by NequIP in
layer 1.  Second, the orbit is complete inside the neighbor list.  Third, orbit members have the same
species and radius.  No assumption is made about the numerical value of $w_p(r_a)$.  This is why the
cancellation is architectural-data interaction rather than a coincidental small learned weight.

\begin{proposition}[Site-symmetry criterion]
The orbit sum $\mathbf S_l^{(a)}$ lies in
$\operatorname{Inv}_{G_i}(D^l\!\downarrow G_i)$, the invariant subspace of the spherical
representation restricted to $G_i$.  Therefore
\begin{equation}
\mathbf1_{G_i}\not\subset D^l\!\downarrow G_i
\quad\Longrightarrow\quad \mathbf S_l^{(a)}=\mathbf0,
\label{eq:trivial_main}
\end{equation}
where $\mathbf1_{G_i}$ is the trivial irrep of the relevant group.
\end{proposition}

For every $g\in G_i$, applying $D^l(g)$ to the sum rotates each spherical harmonic and merely
permutes the members of $\mathcal O_a$.  Thus
$D^l(g)\mathbf S_l^{(a)}=\mathbf S_l^{(a)}$.  Equivalently, the Reynolds projector
$|G_i|^{-1}\sum_{g\in G_i}D^l(g)$ fixes the sum.  If the restricted representation contains no
trivial copy, the projector has no invariant image and the sum must vanish.  The implication is
one-way: even if a trivial component is present, the coefficient of that component can be zero for
a particular orbit geometry.  These restriction and projection arguments follow standard finite-group
representation theory \citep{tung1985group}.  Appendix~\ref{app:proofs} gives the complete derivation.

\begin{figure}[t]
  \centering
  \includegraphics[width=0.94\linewidth]{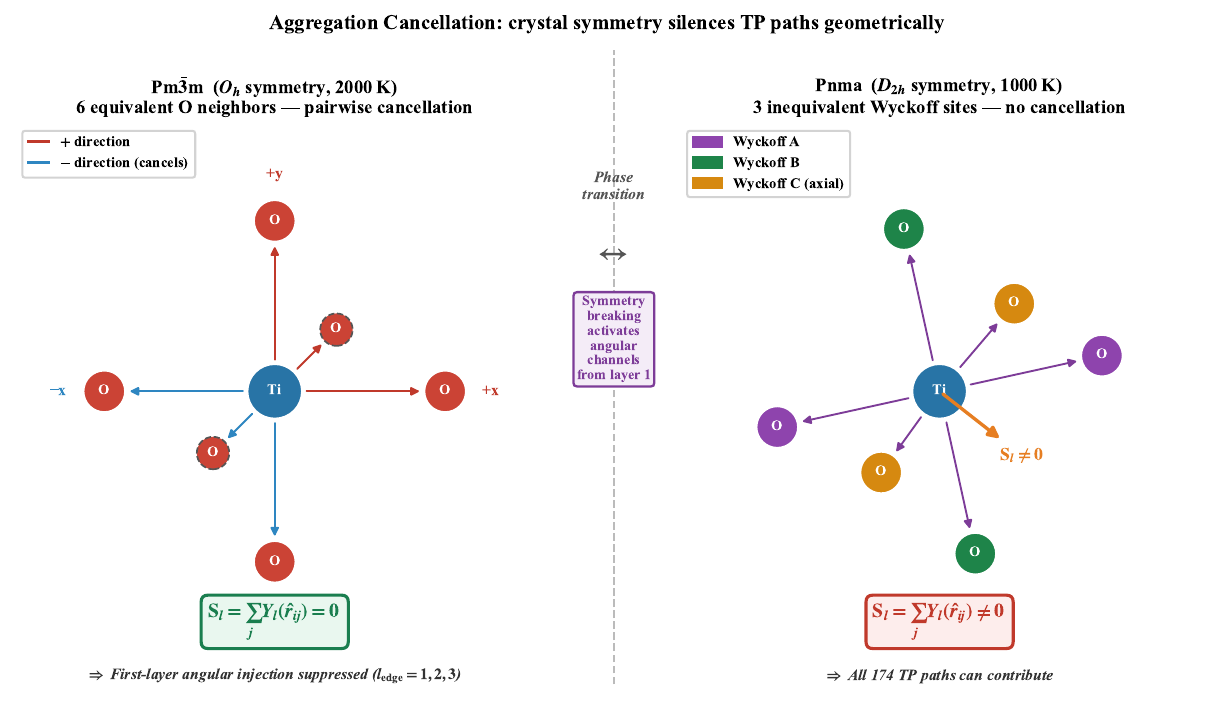}
  \caption{Aggregation Cancellation at an octahedral Ti site.  Six equivalent oxygen neighbors
  form one $O_h$ orbit and their spherical-harmonic sums vanish for $l=1,2,3$.  A lower-symmetry
  distortion partitions the shell into smaller orbits and can activate previously forbidden even
  components.  Exact odd-$l$ cancellation survives whenever an orbit retains inversion-related
  pairs; thermal disorder, non-centrosymmetric sites, or deeper features can relax it.}
  \label{fig:mechanism}
\end{figure}

\subsection{Concrete Site Groups}

For $O_h$, the trivial irrep is conventionally named $A_{1g}$.  The restrictions through $l=3$ are
\begin{align}
D^0\!\downarrow O_h &= A_{1g}, &
D^1\!\downarrow O_h &= T_{1u},\nonumber\\
D^2\!\downarrow O_h &= E_g\oplus T_{2g}, &
D^3\!\downarrow O_h &= A_{2u}\oplus T_{1u}\oplus T_{2u}.
\label{eq:oh_decomp_main}
\end{align}
Only $l=0$ contains $A_{1g}$, so an ideal six-neighbor octahedral orbit transmits only the scalar
first-layer component for $l\le3$.  For a general group we retain the notation
$\mathbf1_{G_i}$: writing $A_{1g}$ for every group would incorrectly assume both the $O_h$ naming
convention and inversion parity.

The tetrahedral group $T_d$ illustrates the distinction between permission and importance.  Its
trivial irrep is $A_1$, and the degree-three representation contains an $A_1$ component.  A
tetrahedral orbit can therefore have a nonzero $l=3$ sum even though its $l=1$ and $l=2$ sums are
forbidden.  Lower-symmetry groups generally admit more invariant components, but the theorem does
not rank those components.  Their amplitudes depend on the orbit geometry and on how the radial and
channel weights respond to chemistry.

Inversion provides a useful special case.  If every direction in an orbit is paired with its
opposite, then $Y_l^m(-\hat{\mathbf r})=(-1)^lY_l^m(\hat{\mathbf r})$, so all odd-$l$ sums cancel.
This remains true for a centrosymmetric distorted orbit even when the full $O_h$ selection rule has
been lost.  It also clarifies why a crystallographic space-group label is only a proxy: a dataset can
contain several Wyckoff sites, thermal snapshots can break instantaneous local symmetry, and a
finite cutoff can truncate an otherwise complete orbit.  Fixed-radius TiO$_6$ probes in trained
CaTiO$_3$ and OMat24 $O_h$ models reproduce ideal cancellation and parity-dependent activation
(Appendix Figure~\ref{fig:distortion_probe_app}); they are not force-accuracy tests.

\subsection{Why the Effect Persists Beyond Layer 1}

The exact proof does not extend mechanically to later layers.  After one interaction,
symmetry-related neighbors satisfy
$h_{g j}^{l}=D^l(g)h_j^l$ rather than sharing one scalar value.  The feature and edge harmonic are
then coupled inside the orbit sum, and the common factor $\alpha_{a,p}$ in
Equation~\eqref{eq:factor_main} cannot generally be extracted.  A direct e3nn enumeration on an
ideal $O_h$ octahedron finds only three active first-layer triplet types but 31 of 34 types can be
active once higher-order neighbor features are supplied.

Nevertheless, the first layer controls the information initially injected into the network.
Directions suppressed there must be reconstructed through distortion, other site types, or deeper
feature interactions.  We therefore expect an upstream symmetry bottleneck to appear statistically
as lower sensitivity of many downstream paths.  The prediction is deliberately qualitative:
near-$O_h$ data should show a lower high-$l$ importance profile than a matched lower-symmetry
condition, while no universal ordering of $l=1,2,3$ follows.  This distinction determines the
experimental design below.

\section{Experimental Design}
\label{sec:experiments}

\subsection{Models, Intervention, and Metrics}

All models use NequIP~0.17.1 with $l_{\max}=3$, five interaction layers, 64 channels per irrep, and
a 5.0\,\AA{} cutoff.  The architecture has 174 weighted layer--instruction blocks.  The edge MLP
emits one contiguous coefficient block for each instruction, so a path can be intervened on by
zeroing the corresponding final-linear columns.  For path $p$, we report
\begin{equation}
\widehat{\Delta\mathrm{MAE}}_p=
\frac{\mathrm{MAE}_{p\text{-ablated}}-\mathrm{MAE}_\mathrm{base}}
     {\mathrm{MAE}_\mathrm{base}}.
\label{eq:normalized_ablation}
\end{equation}
This measures functional sensitivity of a trained model; because the dense instruction list and
tensors are unchanged, ablation alone does not reduce runtime.

The CaTiO$_3$ comparison is the most controlled symmetry experiment.  One model is trained jointly
on 603 MatterSim-driven MD frames \citep{yang2024mattersim} sampled at 1000, 1500, and 2000\,K.  The full split contains
482/60/61 train/validation/test frames.  The ablation analysis contrasts the 1000 and 2000\,K
endpoints, avoiding a comparison between separately trained networks.  The 1000\,K structures are
predominantly distorted Pnma environments, whereas the 2000\,K trajectory is closer to cubic
Pm$\bar3$m.  The temperature label is not treated as an exact site-symmetry annotation.

OMat24 \citep{barroso2024omat24} supplies broader but observational evidence.  We train two
50k-frame perovskite models stratified by temperature and four models stratified by equilibrium
space-group class: $O_h$-cubic, $T_d$-cubic, mid-symmetry, and low-symmetry.  The class label is a
dataset-level proxy and covaries with chemistry.  We therefore use OMat24 to test whether the
predicted trend recurs across data regimes, not as a clean causal isolation of point group.

Two additional datasets provide diagnostic contrasts.  MPF denotes the MPF2021 relaxation corpus
released with M3GNet \citep{chen2022m3gnet}; our F--Fe--O subset contains 372 materials and 1116
density functional theory (DFT) relaxation frames.  Aspirin from MD-17
\citep{chmiela2017md17} is a low-symmetry molecular case.
These datasets contextualise path sensitivity but are not included in the three-system confirmatory
pruning claim.  Their construction and split limitations are described in
Appendix~\ref{app:experiments}.

\subsection{Pruning Evaluation Protocol}

The pruning evaluation assigns a single statistical role to each split.  Training data estimate a
path ranking, validation data select a pruning fraction, and held-out test data estimate the final
accuracy change.  For every weighted path $p$ with edge-MLP block $W_p$, we accumulate force-loss
gradients over $B=10$ training mini-batches and define
\begin{equation}
I_p=\left\|\frac{1}{B}\sum_{b=1}^{B}
\frac{\partial\mathcal L_F^{(b)}}{\partial W_p}\right\|_F.
\label{eq:gradient_score}
\end{equation}
Paths are sorted in ascending $I_p$.  For each cumulative fraction
$q\in\{0,5,10,15,20,25,30,40,50,60\}\%$, the unmodified checkpoint is restored before applying
the corresponding mask.  We select the largest prefix-safe fraction: every evaluated fraction up
to it must remain within the predeclared 5\% validation force-MAE tolerance.  The selected mask is
then frozen and used for one baseline/pruned comparison on the test split.  There is no mask
fine-tuning or test-guided choice in this primary evaluation.  Later controls reuse the inspected
test sets and are not a fresh confirmatory sample.

For CaTiO$_3$, ranking uses 10 balanced endpoint frames drawn from 327 endpoint training frames;
selection uses all 36 endpoint validation frames; and testing uses all 39 endpoint test frames
(23 at 1000\,K and 16 at 2000\,K).  For each OMat24 model, ranking uses 10 frames from 40,000
training frames, selection uses 200 of 5,000 validation frames, and testing uses all 5,000 held-out
frames.  File-level audits find no exactly duplicated frame across the OMat24 splits.  The available
files do not contain a material or trajectory identifier, so these OMat24 splits are accurately
described as frame-disjoint, not proven material-disjoint.

\section{Results}
\label{sec:results}

\subsection{Controlled CaTiO$_3$ Importance Shift}

\begin{figure}[H]
  \centering
  \includegraphics[width=0.82\linewidth]{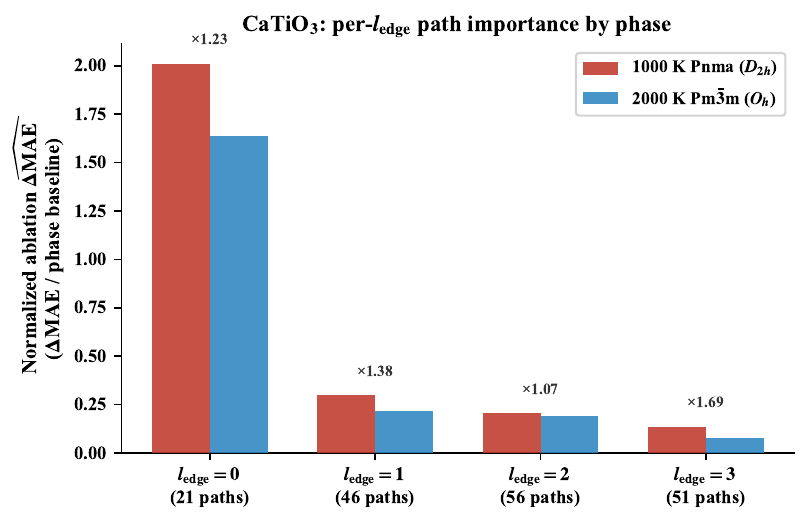}
  \caption{CaTiO$_3$ normalized single-path ablation grouped by edge angular momentum at the 1000
  and 2000\,K endpoints.  One jointly trained model is evaluated under both conditions.  The
  consistent increase for $l_\mathrm{edge}\ge1$ in the lower-symmetry endpoint is the controlled
  test of the predicted importance bias; the theory does not predict an ordering among $l=1,2,3$.}
  \label{fig:catio3_main}
\end{figure}

Figure~\ref{fig:catio3_main} groups the 174 single-path ablations by
$l_\mathrm{edge}$.  Relative to the near-cubic 2000\,K endpoint, the lower-symmetry 1000\,K endpoint
has larger mean normalized ablation for every $l_\mathrm{edge}\ge1$, with endpoint ratios
$r_1=1.38$, $r_2=1.07$, and $r_3=1.69$.  The result matches the predicted direction: angular paths
suppressed by an ideal octahedral orbit become more useful when distortions and heterogeneous local
environments increase.  We do not interpret the ordering of the three ratios.  It is affected by
finite-temperature symmetry breaking, contributions from Ca and O centers, and paths in layers
where exact factorisation no longer applies.

The gradient and intervention rankings agree strongly, with Spearman correlations of $0.85$ at
1000\,K and $0.82$ at 2000\,K.  This supports the use of accumulated gradients as a low-cost mask
ranking signal.  It does not make the ranking data-independent: the score is still calibrated on a
specific distribution, which is why the pruning experiment keeps selection and test structures
separate.

\subsection{Symmetry Permission Does Not Fix Magnitude}

\begin{figure}[t]
  \centering
  \includegraphics[width=0.94\linewidth]{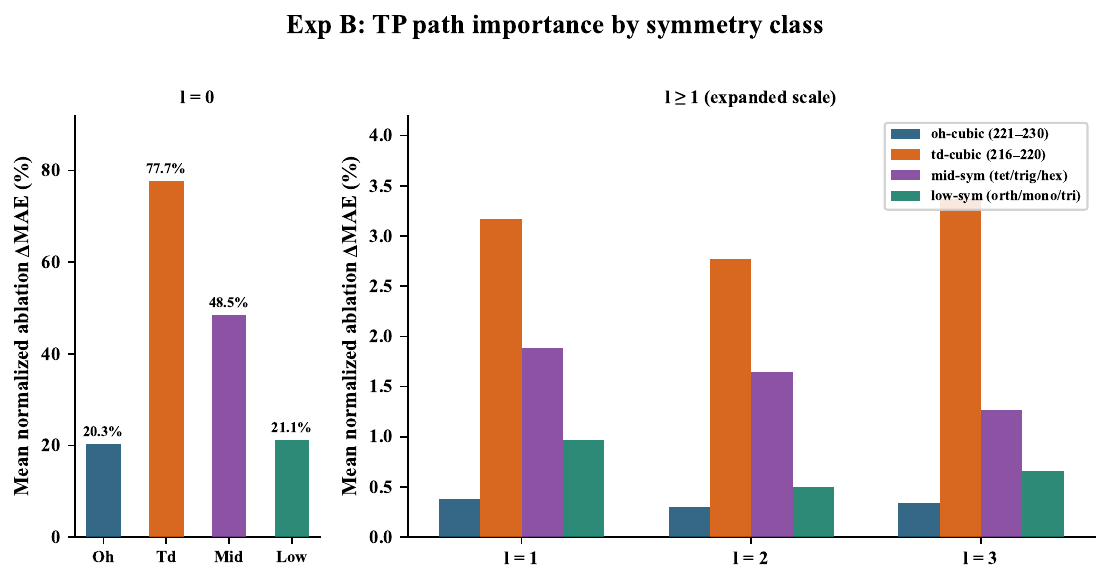}
  \caption{OMat24 mean normalized ablation for models stratified by equilibrium space-group class.
  The $O_h$-cubic subset has the lowest high-$l$ importance, while the $T_d$ subset demonstrates
  that symmetry permission and learned magnitude are different questions.  The full-width panel is
  shown separately from Figure~\ref{fig:catio3_main} to preserve readable labels and uncertainty
  bars.}
  \label{fig:omat_symmetry_main}
\end{figure}

Figure~\ref{fig:omat_symmetry_main} separates the OMat24 space-group models.  The $O_h$-cubic model
has the smallest mean normalized ablation for $l=1,2,3$ ($0.38\%$, $0.30\%$, and $0.34\%$),
consistent with a strong first-layer octahedral bottleneck.  The $T_d$-cubic model instead reaches
$3.17\%$, $2.77\%$, and $3.36\%$.  In particular, $l=3$ is symmetry-allowed for a tetrahedral
orbit, so its large learned importance is compatible with the theorem.  Directional covalent
bonding offers a chemical explanation for why an allowed channel is strongly used.

The mid- and low-symmetry models lie between these extremes for the reported averages.  This is a
useful negative lesson: a lower group permits more invariant components but need not make all of
them more important than in every higher-symmetry chemistry.  Because the four models differ in
composition and training distribution, the comparison is observational.  The evidence is strongest
when read together with the within-model CaTiO$_3$ endpoint shift rather than as a standalone causal
ranking of space groups.

\subsection{Held-Out Pruning Robustness}

\begin{figure}[t]
  \centering
  \includegraphics[width=0.90\linewidth]{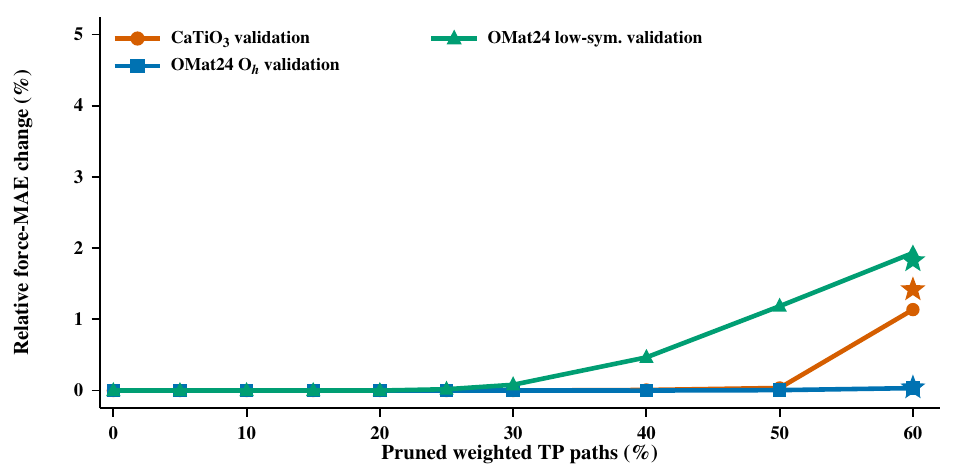}
  \caption{Validation selection curves for the three confirmatory systems.  Lines use validation
  frames only; stars show the held-out test changes after the 60\% masks were frozen.  All
  validation curves remain below the predeclared 5\% tolerance.  Sixty percent is the highest
  evaluated pruning level, not a claimed global maximum.}
  \label{fig:leakfree_curves}
\end{figure}

All three validation curves remain below the 5\% tolerance through the highest evaluated fraction,
60\% (Figure~\ref{fig:leakfree_curves}).  After freezing those masks, held-out force MAE changes by
$+1.42\%$ for CaTiO$_3$, $+0.04\%$ for OMat24 $O_h$-cubic, and $+1.83\%$ for OMat24
low-symmetry (Table~\ref{tab:leakfree_results}).  The corresponding absolute force MAEs are
23.68 to 24.02, 229.19 to 229.28, and 167.22 to 170.27\,meV/\AA{}, respectively.  For CaTiO$_3$,
the held-out changes are also small when separated by endpoint: $+1.80\%$ at 1000\,K and
$+1.13\%$ at 2000\,K.

The $O_h$ result is nearly invariant to the mask, whereas the low-symmetry result has the largest
degradation.  This ordering is consistent with the proposed mechanism, but it is not by itself a
controlled symmetry intervention because the two OMat24 models differ in data distribution.  More
importantly, both remain below the tolerance after the ranking, fraction choice, and test evaluation
are separated.  The result therefore supports pruning robustness under the observed splits rather
than merely reproducing a mask on the data used to choose it.

At fixed 60\% pruning, three OMat24 gradient rankings change test MAE by 0.040--0.065\% ($O_h$)
and 1.825--2.241\% (low symmetry); every layer-matched random mask degrades more (Appendix
Figure~\ref{fig:pruning_controls_app}).  This post hoc result does not isolate symmetry as the cause.

\begin{table}[t]
  \centering
  \small
  \caption{Held-out evaluation at the validation-selected 60\% path mask.  Frame counts
  are ranking/selection/test.  Force MAEs are in meV/\AA{}.  TP and MLP columns are potential
  operation reductions, not measured latency.}
  \label{tab:leakfree_results}
  \setlength{\tabcolsep}{3.3pt}
  \begin{tabular}{lrrrrrr}
    \toprule
    System & Frames & Base & Pruned & $\Delta$ (\%) & TP (\%) & MLP (\%) \\
    \midrule
    CaTiO$_3$ endpoints & 10/36/39 & 23.68 & 24.02 & 1.42 & 71.94 & 59.77 \\
    OMat24 $O_h$ & 10/200/5000 & 229.19 & 229.28 & 0.04 & 67.12 & 59.77 \\
    OMat24 low-sym. & 10/200/5000 & 167.22 & 170.27 & 1.83 & 68.45 & 59.77 \\
    \bottomrule
  \end{tabular}
\end{table}

The selected masks share substantial structure.  The $O_h$ and low-symmetry masks overlap on 89 of
104 paths (Jaccard index 0.75), while CaTiO$_3$ overlaps with them at 0.81 and 0.87 Jaccard,
respectively.  None of the three 60\% masks removes an instruction from the zero-indexed first
interaction layer; the pruned paths are concentrated in later layers.  For example, the
CaTiO$_3$ mask removes 11, 41, and 52 paths from layers 2--4 (zero-based layers 1--3), and the
low-symmetry mask removes 7, 40, and 57.  This does not contradict the theorem.  Exact cancellation
creates an upstream information bottleneck, while gradient ranking identifies the downstream blocks
whose removal least changes the trained computation.  The masks should therefore be understood as
mostly shared downstream sparse subnetworks compatible with, but not uniquely determined by, the
first-layer selection rule.

\subsection{Potential Operation Reduction}

Pruning a fraction of paths does not translate linearly to TP arithmetic because instructions have
different representation dimensions.  For path
$p=(l_\mathrm{in},l_\mathrm{edge},l_\mathrm{out})$ with multiplicity $C_p$, define the
dimension-weighted dense-contraction proxy
\begin{equation}
\phi_p=C_p(2l_\mathrm{in}+1)(2l_\mathrm{edge}+1)(2l_\mathrm{out}+1),\qquad
R_\mathrm{TP}=\frac{\sum_{p\in\mathcal P}\phi_p}{\sum_p\phi_p}.
\label{eq:ops_main}
\end{equation}
The proxy counts products over representation components before implementation-specific
Clebsch--Gordan sparsity.  Because the masks preferentially select some high-dimensional paths,
removing 60\% of the instructions corresponds to 67.12--71.94\% of this potential TP work.

For the final edge-MLP linear layer, each 64-wide path block has the same output width in these
models.  Its removable multiply--accumulate fraction is
\begin{equation}
R_\mathrm{MLP}=\frac{\sum_{p\in\mathcal P}H_{n(p)}C_p}
                         {\sum_pH_{n(p)}C_p}=59.77\%,
\label{eq:mlp_ops_main}
\end{equation}
where $H_{n(p)}$ is the input width of the layer-specific final linear map.  This component-level
accounting deliberately omits scatter, radial basis evaluation, nonlinearities, memory traffic,
force backpropagation, kernel launch overhead, and any CG sparsity exploited by e3nn.  Moreover,
setting columns to zero still executes the original dense graph.  Realizing the potential reduction
requires deleting TP instructions, compacting the edge-MLP outputs, rebuilding the model, and
benchmarking energy-and-force inference on the target hardware.  We therefore use the phrase
\emph{pruning robustness and potential operation reduction}, not compression speedup.

\section{Discussion}
\label{sec:discussion}

\paragraph{Scope and Limitations.}
Aggregation Cancellation connects local site-symmetry orbits to the internal use of an equivariant
network: architectural equivariance permits a coupling, whereas orbit aggregation can annihilate its
actual input.  The criterion is exact only for a complete first-layer orbit satisfying the stated
equal-species and equal-radius assumptions.  It does not rank symmetry-allowed channels, and thermal
disorder, heterogeneous sites, cutoff truncation, and deeper features turn the selection rule into an
empirical importance bias.  All trained-model evidence in this work uses NequIP.  Although the
first-layer factorisation also applies to architectures that share the assumed scalar input and
radial-weighted neighbor aggregation, we do not empirically establish the same importance or pruning
behavior for MACE, Allegro, or other equivariant interatomic potentials.

\paragraph{Design and evaluation implications.}
Persistently symmetric domains may remove provably forbidden first-layer instructions before
training; mixed phases are better served by validation-calibrated masks or conditional gates.  The
observed mask overlap suggests a partially reusable sparse backbone, but transfer requires fresh
validation.  Three post hoc ranking draws show stability at fixed 60\% pruning; OMat24 evaluation
remains frame-disjoint rather than material-disjoint.  The operation proxies identify removable
arithmetic, while acceleration requires structural compilation and hardware timing.

\section{Conclusion}
\label{sec:conclusion}

Within the scalar-input first layer studied here, equivariance determines which angular couplings are
permitted, but the deployment data determine which of those couplings are activated.  We formalised
this distinction through
Aggregation Cancellation: equal-weight neighbors on a site-symmetry orbit project the first-layer
spherical signal onto the trivial subspace, exactly eliminating a TP path when that subspace is
absent.  Because deeper equivariant features break the scalar factorisation, the exact first-layer
rule becomes a network-wide importance bias rather than permanent inactivity.

Matched-radius probes reproduce the predicted first-layer cancellation and distortion response.
Across the NequIP models evaluated here, CaTiO$_3$ ablations and six OMat24 models support the
broader importance bias while showing that
symmetry permission does not determine learned magnitude.  In the primary study, the original 60\%
masks change held-out force MAE by 0.04--1.83\% and imply 67.12--71.94\% potential TP-operation
reduction.  Post hoc OMat24 controls show that gradient masks are stable across ranking seeds and
outperform the random and magnitude-based controls.  These findings support pruning robustness, not
measured compression or speedup.
Material-disjoint evaluation, compiled benchmarks, and validation on architectures such as MACE and
Allegro remain necessary before making a broader claim about equivariant interatomic potentials.

\clearpage
\section*{AI Use Statement}
Generative AI tools were used to critique and refine the presentation of
theoretical material, provide feedback on experimental design, assist with code
implementation and scientific figures, support the interpretation of experimental
results, and help organize, draft, and edit the manuscript and references.  All
AI-assisted content was reviewed and verified by the authors.  All reported
numerical results were obtained from the authors' computational experiments.
The authors take full responsibility for all mathematical claims, code, citations,
experimental results, and final content.

\bibliography{references}

@article{thomas2018tensor,
  title   = {Tensor field networks: Rotation- and translation-equivariant neural networks for 3D point clouds},
  author  = {Thomas, Nathaniel and Smidt, Tess and Kearnes, Steven and Yang, Lusann and Li, Li and Kohlhoff, Kai and Riley, Patrick},
  journal = {arXiv preprint arXiv:1802.08219},
  year    = {2018}
}

@inproceedings{anderson2019cormorant,
  title     = {Cormorant: Covariant molecular neural networks},
  author    = {Anderson, Brandon and Hy, Truong Son and Kondor, Risi},
  booktitle = {Advances in Neural Information Processing Systems},
  volume    = {32},
  year      = {2019}
}

@inproceedings{molchanov2019importance,
  title     = {Importance Estimation for Neural Network Pruning},
  author    = {Molchanov, Pavlo and Mallya, Arun and Tyree, Stephen and Frosio, Iuri and Kautz, Jan},
  booktitle = {Proceedings of the IEEE/CVF Conference on Computer Vision and Pattern Recognition},
  pages     = {11264--11272},
  year      = {2019}
}

@inproceedings{cohen2016group,
  title     = {Group Equivariant Convolutional Networks},
  author    = {Cohen, Taco and Welling, Max},
  booktitle = {International Conference on Machine Learning},
  pages     = {2990--2999},
  year      = {2016}
}

@inproceedings{kondor2018generalization,
  title     = {On the generalization of equivariance and convolution in neural networks to the action of compact groups},
  author    = {Kondor, Risi and Trivedi, Shubhendu},
  booktitle = {International Conference on Machine Learning},
  pages     = {2747--2755},
  year      = {2018}
}

@book{wigner1959theory,
  title     = {Group Theory and its Application to the Quantum Mechanics of Atomic Spectra},
  author    = {Wigner, Eugene P.},
  publisher = {Academic Press},
  year      = {1959}
}

@article{batzner2022nequip,
  title     = {E(3)-equivariant graph neural networks for data-efficient and accurate interatomic potentials},
  author    = {Batzner, Simon and Musaelian, Albert and Sun, Lixin and Geiger, Mario and Mailoa, Jonathan P. and Kornbluth, Mordechai and Molinari, Nicola and Smidt, Tess E. and Kozinsky, Boris},
  journal   = {Nature Communications},
  volume    = {13},
  pages     = {2453},
  year      = {2022},
  publisher = {Nature Publishing Group},
  doi       = {10.1038/s41467-022-29939-5}
}

@inproceedings{batatia2022mace,
  title     = {{MACE}: Higher Order Equivariant Message Passing Neural Networks for Fast and Accurate Force Fields},
  author    = {Batatia, Ilyes and Kov{\'a}cs, D{\'a}vid P{\'e}ter and Simm, Gregor N. C. and Ortner, Christoph and Cs{\'a}nyi, G{\'a}bor},
  booktitle = {Advances in Neural Information Processing Systems},
  volume    = {35},
  year      = {2022}
}

@article{musaelian2023allegro,
  title     = {Learning local equivariant representations for large-scale atomistic dynamics},
  author    = {Musaelian, Albert and Batzner, Simon and Johansson, Anders and Sun, Lixin and Owen, Cameron J. and Kornbluth, Mordechai and Kozinsky, Boris},
  journal   = {Nature Communications},
  volume    = {14},
  pages     = {579},
  year      = {2023},
  doi       = {10.1038/s41467-023-36329-y}
}

@article{barroso2024omat24,
  title   = {{Open Materials 2024 (OMat24)} Inorganic Materials Dataset and Models},
  author  = {Barroso-Luque, Luis and Muhammed, Muhammed Shuaibi and Fu, Xiang and Wood, Brandon M. and Dzamba, Misko and Guo, Jingyu and Schmid, Matthieu and Zitnick, C. Lawrence and Ulissi, Zachary W.},
  journal = {arXiv preprint arXiv:2410.12771},
  year    = {2024}
}

@article{geiger2022e3nn,
  title   = {e3nn: Euclidean Neural Networks},
  author  = {Geiger, Mario and Smidt, Tess},
  journal = {arXiv preprint arXiv:2207.09453},
  year    = {2022}
}

@book{tung1985group,
  title     = {Group Theory in Physics},
  author    = {Tung, Wu-Ki},
  publisher = {World Scientific},
  year      = {1985}
}

@article{chmiela2017md17,
  title     = {Machine learning of accurate energy-conserving molecular force fields},
  author    = {Chmiela, Stefan and Tkatchenko, Alexandre and Sauceda, Huziel E. and Poltavsky, Igor and Sch{\"u}tt, Kristof T. and M{\"u}ller, Klaus-Robert},
  journal   = {Science Advances},
  volume    = {3},
  number    = {5},
  pages     = {e1603015},
  year      = {2017},
  doi       = {10.1126/sciadv.1603015}
}

@inproceedings{lecun1990obd,
  title     = {Optimal Brain Damage},
  author    = {LeCun, Yann and Denker, John and Solla, Sara},
  booktitle = {Advances in Neural Information Processing Systems},
  volume    = {2},
  year      = {1989}
}

@inproceedings{frankle2019lottery,
  title     = {The Lottery Ticket Hypothesis: Finding Sparse, Trainable Neural Networks},
  author    = {Frankle, Jonathan and Carlin, Michael},
  booktitle = {International Conference on Learning Representations},
  year      = {2019}
}

@article{yang2024mattersim,
  title   = {MatterSim: A Deep Learning Atomistic Model Across Elements, Temperatures and Pressures},
  author  = {Yang, Han and Hu, Chenxi and Zhou, Yichi and Liu, Xixian and Shi, Yu and Li, Jielan and Li, Guanzhi and Chen, Zekun and Chen, Shuizhou and Zeni, Claudio and others},
  journal = {arXiv preprint arXiv:2405.04967},
  year    = {2024}
}

@book{varshalovich1988quantum,
  title     = {Quantum Theory of Angular Momentum},
  author    = {Varshalovich, D. A. and Moskalev, A. N. and Khersonskii, V. K.},
  publisher = {World Scientific},
  year      = {1988}
}

@article{park2024sevennet,
  title   = {{SevenNet}: A Interatomic Potential Benchmark for Bulk Phase Materials},
  author  = {Park, Yutack and Kim, Jaesun and Hwang, Seungwoo and Han, Seungwu},
  journal = {arXiv preprint arXiv:2402.03789},
  year    = {2024}
}

@inproceedings{liao2023equiformer,
  title     = {{EquiformerV2}: Improved Equivariant Transformer for Scaling to Higher-Degree Representations},
  author    = {Liao, Yi-Lun and Wood, Brandon and Das, Abhishek and Smidt, Tess},
  booktitle = {International Conference on Learning Representations},
  year      = {2024}
}

@article{drautz2019ace,
  title     = {Atomic cluster expansion for accurate and transferable interatomic potentials},
  author    = {Drautz, Ralf},
  journal   = {Physical Review B},
  volume    = {99},
  number    = {1},
  pages     = {014104},
  year      = {2019},
  doi       = {10.1103/PhysRevB.99.014104}
}

@article{pozdnyakov2022incompleteness,
  title     = {Incompleteness of Atomic Structure Representations},
  author    = {Pozdnyakov, Sergey N. and Willatt, Michael J. and Bart{\'o}k, Albert P. and Ortner, Christoph and Cs{\'a}nyi, G{\'a}bor and Ceriotti, Michele},
  journal   = {Physical Review Letters},
  volume    = {125},
  number    = {16},
  pages     = {166001},
  year      = {2020},
  doi       = {10.1103/PhysRevLett.125.166001}
}

@article{nigam2022unified,
  title     = {Unified theory of atom-centered representations and message-passing machine-learning schemes},
  author    = {Nigam, Jigyasa and Pozdnyakov, Sergey and Cs{\'a}nyi, G{\'a}bor and Ceriotti, Michele},
  journal   = {The Journal of Chemical Physics},
  volume    = {156},
  number    = {20},
  pages     = {204115},
  year      = {2022},
  doi       = {10.1063/5.0087042}
}

@inproceedings{passaro2023reducing,
  title     = {Reducing {SO(3)} Convolutions to {SO(2)} for Efficient Equivariant {GNN}s},
  author    = {Passaro, Saro and Zitnick, C. Lawrence},
  booktitle = {International Conference on Machine Learning},
  pages     = {27420--27438},
  year      = {2023}
}

@inproceedings{xie2025price,
  title     = {The Price of Freedom: Exploring Expressivity and Runtime Tradeoffs in Equivariant Tensor Products},
  author    = {Xie, Yuqing and Daigavane, Ameya and Kotak, Mit and Smidt, Tess E.},
  booktitle = {Proceedings of the 42nd International Conference on Machine Learning},
  series    = {Proceedings of Machine Learning Research},
  volume    = {267},
  pages     = {68599--68625},
  year      = {2025}
}

@inproceedings{lee2025flashtp,
  title     = {{FlashTP}: Fused, Sparsity-Aware Tensor Product for Machine Learning Interatomic Potentials},
  author    = {Lee, Seung Yul and Kim, Hojoon and Park, Yutack and Jeong, Dawoon and Han, Seungwu and Park, Yeonhong and Lee, Jae W.},
  booktitle = {Proceedings of the 42nd International Conference on Machine Learning},
  series    = {Proceedings of Machine Learning Research},
  volume    = {267},
  pages     = {33143--33156},
  year      = {2025}
}

@inproceedings{luo2025sphnet,
  title     = {Efficient and Scalable Density Functional Theory Hamiltonian Prediction through Adaptive Sparsity},
  author    = {Luo, Erpai and Wei, Xinran and Huang, Lin and Li, Yunyang and Yang, Han and Xia, Zaishuo and Wang, Zun and Liu, Chang and Shao, Bin and Zhang, Jia},
  booktitle = {Proceedings of the 42nd International Conference on Machine Learning},
  series    = {Proceedings of Machine Learning Research},
  volume    = {267},
  pages     = {41368--41390},
  year      = {2025}
}

@article{heilman2026pointgroup,
  title   = {Point Group Equivariant Graph Neural Networks for Materials},
  author  = {Heilman, Alexander J. and Yan, Qimin},
  journal = {arXiv preprint arXiv:2607.16871},
  year    = {2026}
}

@article{chen2022m3gnet,
  title   = {A Universal Graph Deep Learning Interatomic Potential for the Periodic Table},
  author  = {Chen, Chi and Ong, Shyue Ping},
  journal = {Nature Computational Science},
  volume  = {2},
  pages   = {718--728},
  year    = {2022}
}

@inproceedings{frank2022so3krates,
  title     = {So3krates: Equivariant attention for interactions on arbitrary length-scales in molecular systems},
  author    = {Frank, J. Thorben and Unke, Oliver T. and M{\"u}ller, Klaus-Robert},
  booktitle = {Advances in Neural Information Processing Systems},
  volume    = {35},
  year      = {2022}
}
\bibliographystyle{iclr2027_conference}

\clearpage
\appendix
\section{Proofs and Group-Theoretic Details}
\label{app:proofs}

\subsection{Orbit Factorisation}

In the first layer, a neighbor $j$ of species $s_a$ carries the scalar feature
$h_j^{0,0}=h_{s_a}$.  For all $j$ in a site-symmetry orbit $\mathcal O_a$, the distance
$r_{ij}=r_a$ and species are constant.  Restricting Equation~\eqref{eq:message_main} to this
orbit and using $m_\text{in}=0$ gives
\begin{align}
M^{l_\text{out}}_{i,m}(\mathcal O_a)
&=w(r_a)h_{s_a}\sum_{j\in\mathcal O_a}
  \sum_{m_\text{edge}}C^m_{0,m_\text{edge}}
  Y_{l_\text{edge}}^{m_\text{edge}}(\hat{\mathbf r}_{ij})\\
&=\alpha_a\,S^{(a)}_{l_\text{edge},m},
\end{align}
where the CG selection rule fixes $m=m_\text{edge}$ for $l_\text{in}=0$.  Thus the learned
radial function enters only through the common scalar $\alpha_a$ and cannot undo a vanishing
angular sum.

\subsection{Invariant-Subspace Criterion}

Let $G_i$ act on the directions in $\mathcal O_a$.  For every $g\in G_i$,
\begin{align}
D^l(g)\mathbf S_l^{(a)}
&=\sum_{j\in\mathcal O_a}D^l(g)Y_l(\hat{\mathbf r}_{ij})
=\sum_{j\in\mathcal O_a}Y_l(g\hat{\mathbf r}_{ij})
=\mathbf S_l^{(a)},
\end{align}
because $g$ permutes orbit members.  Hence $\mathbf S_l^{(a)}$ belongs to
$\operatorname{Inv}_{G_i}(D^l|_{G_i})$.  Equivalently, the Reynolds projector
$P_{G_i}^{(l)}=|G_i|^{-1}\sum_{g\in G_i}D^l(g)$ satisfies
$P_{G_i}^{(l)}\mathbf S_l^{(a)}=\mathbf S_l^{(a)}$.  If the trivial representation
$\mathbf1_{G_i}$ is absent from the restricted representation, this projector is zero and
$\mathbf S_l^{(a)}=\mathbf0$.

For a centrosymmetric orbit, every direction has an opposite partner and
$Y_l^m(-\hat{\mathbf r})=(-1)^lY_l^m(\hat{\mathbf r})$; all odd-$l$ sums therefore vanish.
This parity argument is a special case of the invariant-subspace criterion.  The notation
$A_{1g}$ is used only for centrosymmetric point groups such as $O_h$; for $T_d$ the trivial
irrep is $A_1$, and $\mathbf1_{G_i}$ is the group-independent notation.

\begin{table}[h]
  \centering
  \caption{Restrictions of $D^l$ to $O_h$ for $l\le3$.}
  \label{tab:oh_decomp_app}
  \begin{tabular}{cclc}
    \toprule
    $l$ & $D^l\downarrow O_h$ & Contains $A_{1g}$? & Ideal orbit sum \\
    \midrule
    0 & $A_{1g}$ & yes & non-zero allowed \\
    1 & $T_{1u}$ & no & zero \\
    2 & $E_g\oplus T_{2g}$ & no & zero \\
    3 & $A_{2u}\oplus T_{1u}\oplus T_{2u}$ & no & zero \\
    \bottomrule
  \end{tabular}
\end{table}

\subsection{Pnma, $T_d$, and Deeper Layers}

Octahedral tilting in Pnma splits the ideal $O_h$ oxygen shell into smaller orbits and can activate
even-$l$ components that $O_h$ forbids.  The ideal Ti site nevertheless remains centrosymmetric,
so opposite-pair cancellation still removes odd $l$ in its first-layer scalar orbit sum.  Odd-$l$
importance measured for the full network can instead arise from thermally broken instantaneous
geometry, non-centrosymmetric atomic sites, or paths in deeper layers.  We therefore do not use
the CaTiO$_3$ $l=3$ ratio as a direct proof of first-layer Ti-site activation.

For a tetrahedral orbit, $T_d$ symmetry forbids $l=1,2$ but permits a trivial component at $l=3$.
Lower-symmetry groups permit progressively more components, although permission alone does not
determine their learned magnitude.  In deeper NequIP layers, symmetry-related neighbors obey
$h_{j'}^{l}=D^l(g)h_j^l$ rather than carrying identical scalars.  A direct e3nn calculation on an
ideal $O_h$ octahedron finds only three non-zero layer-1 triplets but 31 of 34 path types active once
higher-order neighbor features are present.

\begin{figure}[h]
  \centering
  \includegraphics[width=0.85\linewidth]{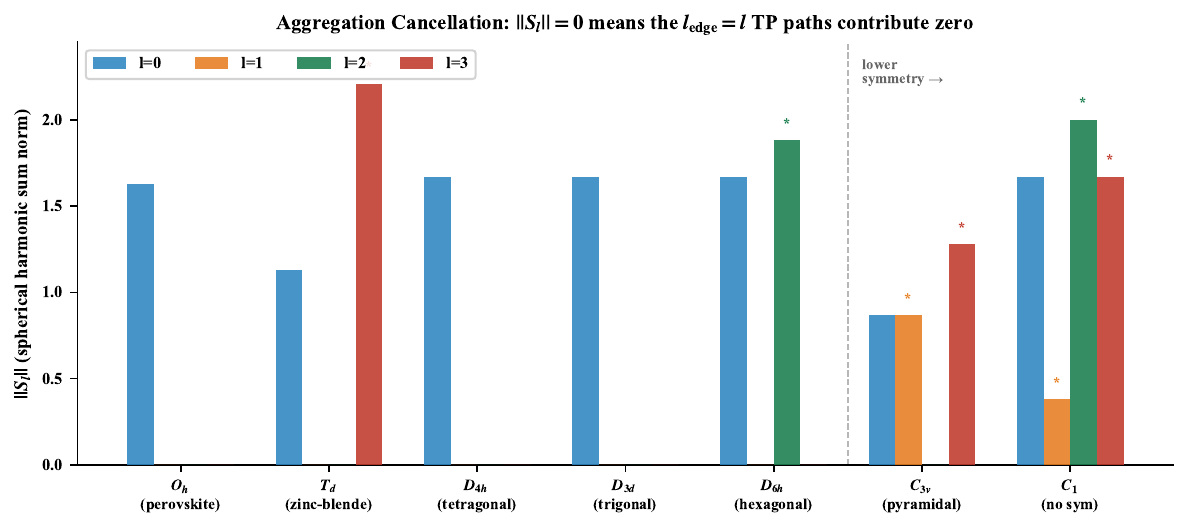}
  \caption{Norms of spherical-harmonic orbit sums for representative site groups.  A zero bar
  indicates a symmetry-forbidden component; an allowed component may still vanish for a particular
  orbit geometry.}
  \label{fig:sl_norms_app}
\end{figure}

\section{Experimental and Dataset Details}
\label{app:experiments}

\subsection{Architecture and Training}

Models use NequIP~0.17.1, five interaction layers, 64 features per irrep, $l_{\max}=3$, and
$r_{\max}=5.0$\,\AA.  Training uses Adam, exponential moving average (EMA) weights with decay 0.999,
and a force-dominated energy--force objective.  All reported evaluations use EMA checkpoints.

\paragraph{CaTiO$_3$.}
MatterSim-driven trajectories contain 603 total frames: 201 each at 1000, 1500, and 2000\,K.  The
482/60/61 train/validation/test files contain temperature counts 163/155/164, 15/24/21, and
23/22/16, respectively.  One model is trained jointly on all three temperatures.  The symmetry and
pruning analyses contrast the 1000 and 2000\,K endpoints; 1500\,K remains part of training but is
excluded from the endpoint evaluation.  This design makes the importance comparison within-model
rather than a comparison between independently trained networks.

\paragraph{OMat24.}
For the temperature study, two models are trained on 50k ABO$_3$ frames at 1000 and 3000\,K.
For the space-group study, structures are partitioned into $O_h$-cubic (space groups 221--230),
$T_d$-cubic (216--220), mid-symmetry (tetragonal, trigonal, and hexagonal), and low-symmetry
(orthorhombic, monoclinic, and triclinic) subsets.  Equilibrium space group is a dataset-level proxy
and does not imply exact instantaneous site symmetry at finite temperature.

\paragraph{MPF2021 F--Fe--O.}
MPF2021 is the DFT relaxation corpus released with M3GNet \citep{chen2022m3gnet}.  We filter
structures whose element set is exactly \{F, Fe, O\}, obtaining 372 materials and 1116 frames.
Using seed 42, the current dataset files contain 780 training, 112 validation, and 224 test frames.
The existing preparation script performs a frame-level split; a material-disjoint split is preferred
for a final generalisation claim and should be regenerated before submission if multiple relaxation
frames share a material identifier across partitions.

\paragraph{Aspirin.}
The molecular contrast uses aspirin from MD-17 \citep{chmiela2017md17}, with 950 training,
50 validation, and 1000 test structures generated using seed 123.  It uses the same five-layer,
$l_{\max}=3$, 64-channel NequIP architecture and is included as a low-symmetry counterexample,
not as an accuracy benchmark against inorganic models.

\subsection{Per-Path Ablation}

For each of the 174 layer--instruction blocks, we restore the original checkpoint, zero only the
associated edge-MLP output columns, and evaluate force MAE.  This intervention measures functional
sensitivity but leaves the dense execution graph unchanged.

\begin{algorithm}[t]
\DontPrintSemicolon
\SetAlgoLined
\caption{Per-Path Ablation for TP Path Importance}
\label{alg:ablation}

\KwIn{Trained NequIP module $\mathcal{M}$; test loader $\mathcal{D}$;
      number of eval batches $B$}
\KwOut{Per-path importance scores $\{\widehat{\DMAE}_k\}_{k=1}^{174}$}

\BlankLine
$\mathrm{entries} \leftarrow []$\;
\For{$n \in \{0,\ldots,4\}$}{
  $\mathrm{conv} \leftarrow \mathcal{M}.\mathrm{layer}_n.\mathrm{conv}$\;
  $\mathrm{tp}  \leftarrow \mathrm{conv}.\mathrm{tp\_scatter}.\mathrm{tp}$\;
  $W \leftarrow \mathrm{conv}.\mathrm{edge\_mlp}.\mathrm{last\_linear}.\mathrm{weight}$\;
  $c \leftarrow 0$\;
  \ForEach{instruction $\iota \in \mathrm{tp.instructions}$}{
    $l_\mathrm{edge} \leftarrow \mathrm{tp.irreps\_in2}[\iota.\mathrm{i\_in2}].\mathrm{ir}.l$\;
    $\mathrm{entries}.\mathrm{append}\!\left(\{W,\; \mathrm{col\_start}{=}c,\; \mathrm{col\_end}{=}c{+}64,\; l_\mathrm{edge}\}\right)$\;
    $c \leftarrow c + 64$\;
  }
}

\BlankLine
$\mathrm{MAE}_\mathrm{base} \leftarrow \mathrm{EvalForceMAE}(\mathcal{M},\, \mathcal{D},\, B)$\;

\BlankLine
$W_\mathrm{orig} \leftarrow \{W.\mathrm{data.clone()}\;\forall\, W\}$\;
\For{$k \leftarrow 1$ \KwTo $|\mathrm{entries}|$}{
  $e \leftarrow \mathrm{entries}[k]$\;
  $e.W.\mathrm{data}[:,\; e.\mathrm{col\_start}\,:\,e.\mathrm{col\_end}]
    \leftarrow \mathbf{0}$\;
  $\mathrm{MAE}_k \leftarrow \mathrm{EvalForceMAE}(\mathcal{M},\, \mathcal{D},\, B)$\;
  \ForEach{$(W, W_0) \in W_\mathrm{orig}$}{$W.\mathrm{data} \leftarrow W_0$}\;
  $\widehat{\DMAE}_k \leftarrow (\mathrm{MAE}_k - \mathrm{MAE}_\mathrm{base})\;/\;\mathrm{MAE}_\mathrm{base}$\;
}
\Return $\{\widehat{\DMAE}_k\}$\;
\end{algorithm}

\section{Pruning Protocol and Operation Counts}
\label{app:pruning}

The implemented protocol uses training data for ranking, validation data for mask selection, and the
held-out test split only after the mask has been frozen.  The complete procedure is given in
Algorithm~\ref{alg:gradient_pruning}.  For CaTiO$_3$, the ranking pool contains
327 endpoint training frames and the ten ranking frames are sampled in a balanced 5/5 split between
1000 and 2000\,K.  All 36 endpoint validation frames are used for selection and all 39 endpoint test
frames are used for the final report.  For both OMat24 models, the ranking sample is 10 of 40,000
training frames, the selection sample is 200 of 5,000 validation frames, and the final evaluation
uses all 5,000 test frames.  The test set is never used to rank paths, choose a threshold, or select
among masks.

\begin{algorithm}[t]
\DontPrintSemicolon
\SetAlgoLined
\caption{Gradient-Ranked Cumulative TP Path Pruning}
\label{alg:gradient_pruning}

\KwIn{Trained model $\mathcal{M}$; training-ranking, validation-selection, and held-out test loaders
      $\mathcal{D}_\text{rank}$, $\mathcal{D}_\text{select}$, $\mathcal{D}_\text{test}$;
      pruning fractions $\mathcal{Q}$; tolerance $\epsilon$; ranking batches $B$}
\KwOut{Frozen mask $\mathcal{P}_{q^*}$; one held-out test record}

$\mathcal{E}\leftarrow$ collect all weighted $(\text{layer},\text{TP instruction})$ entries;
\ForEach{$p\in\mathcal{E}$}{
  locate its edge-MLP column block $\mathbf{W}_p$;
  compute on $\mathcal{D}_\text{rank}$:
  $I_p\leftarrow\left\|B^{-1}\sum_{b=1}^{B}
    \partial\mathcal{L}^{(b)}_F/\partial\mathbf{W}_p\right\|_F$;
  compute $\phi_p\leftarrow C_p(2l_\text{in}{+}1)(2l_\text{edge}{+}1)(2l_\text{out}{+}1)$;
}
$\pi\leftarrow\operatorname{argsort}_{p\in\mathcal{E}} I_p$ in ascending order;
$\Theta_0\leftarrow$ clone original edge-MLP weights;

${\rm MAE}^{\rm select}_0\leftarrow
  \operatorname{EvalForceMAE}(\mathcal M,\mathcal D_\text{select})$;
\ForEach{$q\in\mathcal{Q}$}{
  restore $\Theta_0$;
  $K_q\leftarrow\operatorname{round}(q|\mathcal{E}|/100)$;
  $\mathcal{P}_q\leftarrow\{\pi_1,\ldots,\pi_{K_q}\}$;
  \ForEach{$p\in\mathcal{P}_q$}{$\mathbf{W}_p\leftarrow\mathbf{0}$}
  ${\rm MAE}^{\rm select}(q)\leftarrow
    \operatorname{EvalForceMAE}(\mathcal{M},\mathcal{D}_\text{select})$;
}
$q^*\leftarrow\max\{q:\forall q'\in\mathcal Q,\ q'\leq q\Rightarrow
  {\rm MAE}^{\rm select}(q')\leq(1+\epsilon){\rm MAE}^{\rm select}_0\}$;
restore $\Theta_0$; apply the frozen mask $\mathcal P_{q^*}$;
evaluate $\mathcal D_\text{test}$ exactly once and compute operation reductions;
restore $\Theta_0$ and return $\mathcal P_{q^*}$ and the test record;
\end{algorithm}

The CaTiO$_3$ manifest records file hashes and a temperature-aware split.  Exact structure hashes
for the OMat24 $O_h$ and low-symmetry exports show zero exact duplicates across every train/validation,
train/test, and validation/test pair.  Stable material and trajectory identifiers are absent,
however, so this verifies frame-level disjointness rather than material-level disjointness.

For a directed edge, define $d_l=2l+1$ and
\begin{equation}
\phi_p=C_p d_{l_\mathrm{in}}d_{l_\mathrm{edge}}d_{l_\mathrm{out}},\qquad
R_\mathrm{TP}=\frac{\sum_{p\in\mathcal P}\phi_p}{\sum_p\phi_p}.
\label{eq:tp_flops_reduction}
\end{equation}
Counting one multiply and one add as two operations changes numerator and denominator by the same
factor.  For the final edge-MLP linear layer with hidden width $H_n$,
\begin{equation}
R_\mathrm{MLP}=\frac{\sum_{p\in\mathcal P}H_{n(p)}C_p}
                         {\sum_pH_{n(p)}C_p}.
\label{eq:mlp_flops_reduction}
\end{equation}
These proxies omit CG sparsity, fused-kernel schedules, nonlinearities, scatter, memory traffic,
and reverse-mode force evaluation.  They quantify removable arithmetic only after structural
instruction deletion and edge-MLP compaction.

\section{Additional Results}
\label{app:results}

\subsection{Matched-Radius First-Layer Distortion Probe}

To test the orbit argument within trained networks, we construct a nonperiodic seven-atom TiO$_6$
cluster with one Ti center and six identical O neighbors at a fixed Ti--O radius of 2.0\,\AA.
The ideal oxygen directions are $\pm\hat x$, $\pm\hat y$, and $\pm\hat z$.  At each angle
$\theta\in\{2^\circ,5^\circ,10^\circ,15^\circ\}$, the inversion-preserving case rotates the $+z$ oxygen by
$\theta$ toward $+x$ and moves its opposite partner to the negative of the new direction.  The
inversion-broken case moves only the $+z$ oxygen.  Species, bond lengths, and the number of incoming
Ti--O edges are unchanged.

\begin{figure}[H]
  \centering
  \includegraphics[width=0.94\linewidth]{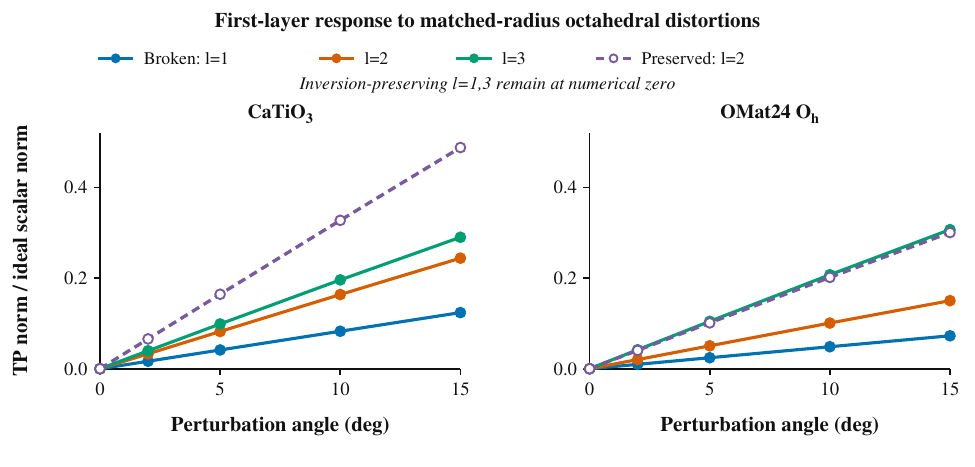}
  \caption{Trained first-layer TP-scatter norms at the Ti center, normalized separately for each
  model by its ideal $l=0$ norm.  Solid lines break inversion; the dashed line preserves opposite
  oxygen pairs.  In the latter case, the $l=1,3$ responses remain at numerical zero and are omitted
  from the plotted series.  All six Ti--O bonds remain at 2.0\,\AA.}
  \label{fig:distortion_probe_app}
\end{figure}

We compute the norm of the degree-$l$ spherical-harmonic sum with e3nn component normalization and
capture the pre-linear output of the trained first-layer TP scatter at the Ti center.  The latter is
grouped by output degree $l$ and divided in Figure~\ref{fig:distortion_probe_app} by the model's
ideal-geometry $l=0$ norm.  For both the CaTiO$_3$ and OMat24 $O_h$ checkpoints, the ideal
$l=1,2,3$ responses are below $3.1\times10^{-8}$ in raw norm.  Under inversion-preserving tilts,
the odd-$l$ responses remain below $2.1\times10^{-8}$ while $l=2$ increases with angle.  Moving
only one oxygen also activates $l=1$ and $l=3$.  The probe isolates geometric symmetry in one
trained layer; it does not evaluate forces or simulate a thermodynamic distortion trajectory.

\subsection{CaTiO$_3$ Ratios}

\begin{table}[H]
  \centering
  \caption{Normalized CaTiO$_3$ ablation ratio between 1000 and 2000\,K.  The selection rule
  predicts a directional bias for symmetry-suppressed channels, not an ordering of the ratios.}
  \begin{tabular}{ccccc}
    \toprule
    $l_\mathrm{edge}$ & 0 & 1 & 2 & 3 \\
    \midrule
    Paths across five layers & 21 & 46 & 56 & 51 \\
    $r_l$ & 1.23 & 1.38 & 1.07 & 1.69 \\
    \bottomrule
  \end{tabular}
\end{table}

\subsection{OMat24 Temperature Study}

\begin{figure}[H]
  \centering
  \includegraphics[width=0.78\linewidth]{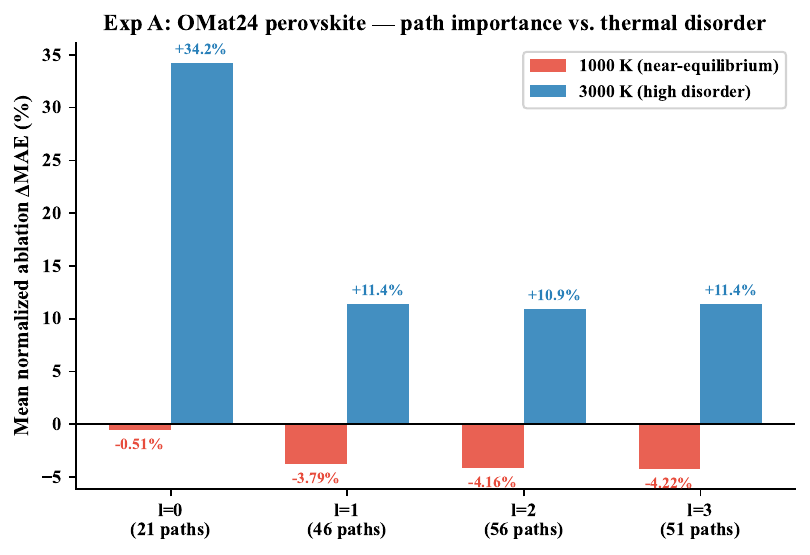}
  \caption{Mean normalized ablation for OMat24 perovskite models.  Near-equilibrium 1000\,K
  structures show negative high-$l$ ablation, whereas thermal disorder at 3000\,K activates the
  same path groups.}
  \label{fig:omat_temp_app}
\end{figure}

\begin{table}[H]
  \centering
  \caption{OMat24 mean normalized ablation $\overline{\widehat{\Delta\mathrm{MAE}}}$ (\%).}
  \begin{tabular}{crrrr}
    \toprule
    $l_\mathrm{edge}$ & $O_h$-cubic & $T_d$-cubic & mid-sym. & low-sym. \\
    \midrule
    0 & 20.3 & 77.7 & 48.5 & 21.1 \\
    1 & 0.38 & 3.17 & 1.88 & 0.96 \\
    2 & 0.30 & 2.77 & 1.64 & 0.50 \\
    3 & 0.34 & 3.36 & 1.26 & 0.66 \\
    \bottomrule
  \end{tabular}
\end{table}

\subsection{Additional Pruning Results}

\begin{table}[H]
  \centering
  \caption{Validation and held-out test changes at the selected 60\% mask.  All values are relative
  force-MAE changes in percent.}
  \label{tab:pruning_validation_app}
  \begin{tabular}{lrrr}
    \toprule
    System & Validation & Test & Test frames \\
    \midrule
    CaTiO$_3$ endpoints & 1.136 & 1.418 & 39 \\
    OMat24 $O_h$-cubic & 0.034 & 0.040 & 5000 \\
    OMat24 low-symmetry & 1.931 & 1.825 & 5000 \\
    \bottomrule
  \end{tabular}
\end{table}

Validation and test changes are close for all three systems, with no evidence that the selected
fraction exploits the validation subset.  The experiment stops at 60\%, so these results establish
robustness at the highest evaluated level but do not locate the failure point.  Earlier same-loader
diagnostics are reported in Appendix~\ref{app:exploratory_pruning} for transparency, but are not
used to support the confirmatory claim.

The 104-path masks are distributed primarily over later interaction layers.  Using zero-based layer
indices, CaTiO$_3$ removes 11/41/52 paths from layers 1/2/3, $O_h$ removes 16/43/45, and
low-symmetry removes 7/40/57; none removes a layer-0 instruction.  The $O_h$/low-symmetry masks
share 89 paths (Jaccard 0.748), while the CaTiO$_3$ mask has Jaccard overlaps 0.809 and 0.874 with
$O_h$ and low-symmetry.  This downstream concentration is compatible with an upstream symmetry
bottleneck affecting sensitivity, but does not by itself identify the cause of the final mask.

\subsection{Fixed-Rate OMat24 Pruning Controls}

We examine the stability of the original OMat24 result at a \emph{fixed} 60\% pruning fraction,
equivalent to 104 of 174 weighted paths.  For each model, three gradient rankings use separately
sampled 10-frame training subsets with seeds 20260910--20260912.  Three unrestricted random masks sample
the same number of paths.  Three layer-matched random masks preserve the first gradient mask's
per-layer path counts, and one magnitude mask removes the paths with the smallest RMS weight in
their final edge-MLP linear blocks.  The same 5,000 test frames are used for every mask within a
model.  No mask is fine-tuned.  These controls were designed after inspection of the original test
result and therefore do not constitute a second independent confirmatory test.

Across gradient seeds, the $O_h$ force-MAE increase is 0.040--0.065\% (mean $0.055\pm0.013\%$),
and the low-symmetry increase is 1.825--2.241\% (mean $2.084\pm0.226\%$).  The $\pm$ values are
sample standard deviations across three masks, not test-set confidence intervals.  Layer-matched
random masks give means of $32.00\pm4.40\%$ and $124.99\pm10.35\%$, respectively.  The
unrestricted random means are $149.67\%$ and $158.28\%$; the single magnitude masks give
$220.54\%$ and $307.48\%$.  These equal-path-count comparisons support the practical value of
the gradient ranking, including beyond its preference for particular layers.  They do not match
the masks by TP-operation proxy or angular composition, and they cannot establish that site
symmetry alone causes the separation.

\begin{figure}[H]
  \centering
  \includegraphics[width=0.94\linewidth]{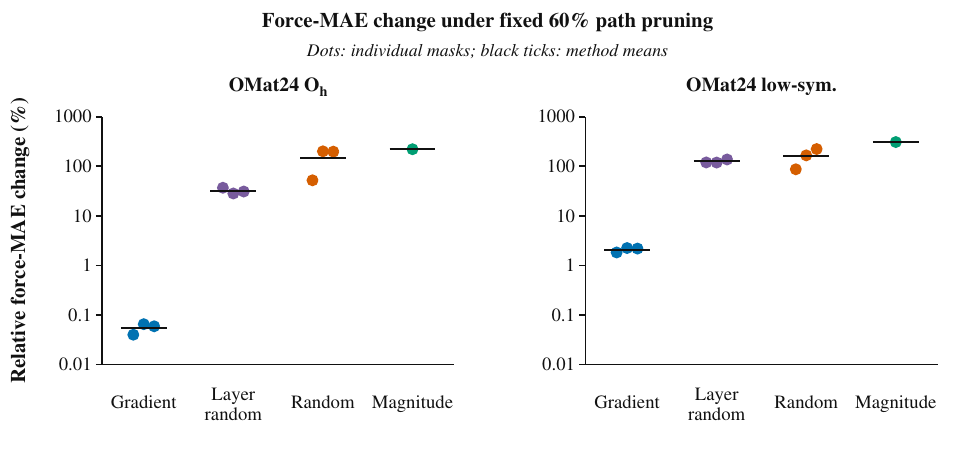}
  \caption{Post hoc OMat24 controls at fixed 60\% path pruning, evaluated on the same 5,000 test
  frames per model.  The vertical axis is logarithmic and shows the percentage increase in force
  MAE relative to the unpruned checkpoint.  Each dot is one mask; black horizontal ticks show
  method means.  Gradient, layer-matched random, and unrestricted random have three seeds each;
  magnitude has one mask.  This is a robustness diagnostic, not a new test split.}
  \label{fig:pruning_controls_app}
\end{figure}

\subsection{Exploratory Same-Loader Pruning Diagnostics}
\label{app:exploratory_pruning}

Before fixing the confirmatory evaluation protocol, we ran diagnostic pruning sweeps on the
OMat24 perovskite models at 1000 and 3000\,K, the MPF2021 F--Fe--O model, and the MD-17 Aspirin
model.  These experiments use the same loader both to estimate gradient importance and to evaluate
the pruned masks.  They are retained to document how the pruning strategy behaves across different
chemical domains, but they do not provide an independent estimate of mask generalisation.

\begin{figure}[H]
  \centering
  \includegraphics[width=0.82\linewidth]{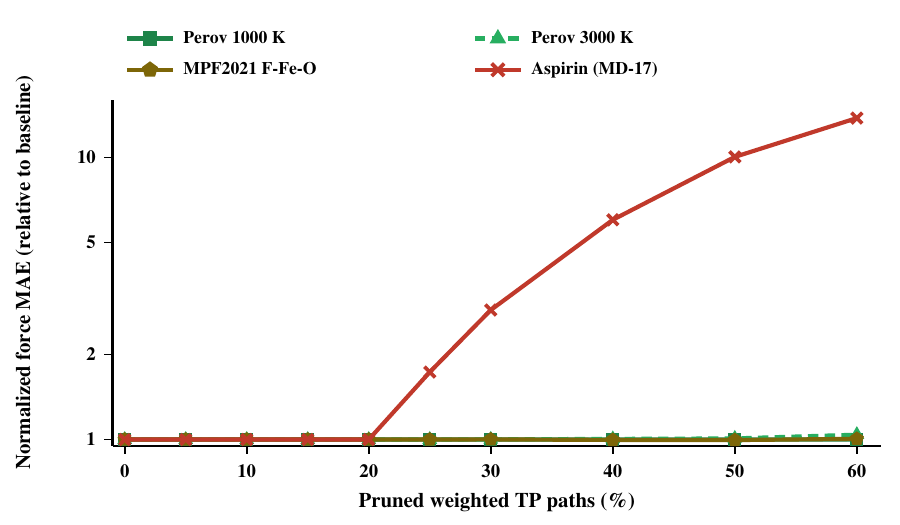}
  \caption{Exploratory same-loader pruning curves, normalized by each model's unpruned force MAE
  and shown on a logarithmic vertical scale.
  At 60\% pruning, the relative changes are 0.02\% for OMat24 perovskite at 1000\,K, 4.05\% at
  3000\,K, and 0.60\% for MPF2021 F--Fe--O.  Aspirin is qualitatively different: degradation begins
  between 20\% and 25\% pruning and reaches $13.81\times$ the baseline MAE at 60\%.  Because ranking
  and evaluation reuse the same loader, these curves are descriptive diagnostics only.}
  \label{fig:exploratory_pruning}
\end{figure}

The descriptive curves are shown in Figure~\ref{fig:exploratory_pruning}.

For each trained $l_{\max}=3$ model, all 174 weighted layer--instruction paths are enumerated.  If
$W_p$ denotes the edge-MLP final-linear block associated with path $p$, its diagnostic score is the
Frobenius norm of the accumulated force-MAE gradient over the first ten mini-batches,
\begin{equation}
I_p^{\mathrm{diag}}=\left\|\sum_{b=1}^{10}
\frac{\partial\mathcal L_F^{(b)}}{\partial W_p}\right\|_F.
\label{eq:diagnostic_gradient_score}
\end{equation}
The paths are sorted in ascending $I_p^{\mathrm{diag}}$ and cumulative prefixes corresponding to
$q\in\{0,5,10,15,20,25,30,40,50,60\}\%$ are zeroed without retraining.  The original checkpoint
weights are restored before every value of $q$.  The intervention zeros the relevant edge-MLP
columns while leaving the dense tensor-product instruction list unchanged, so the reported
operation reductions remain structural proxies rather than measured speedups.

\begin{table}[H]
  \centering
  \small
  \caption{Data roles in the exploratory pruning sweeps.  ``Same loader'' means that the data used
  to form the gradient ranking are also used to compute the pruning curve.  All rankings use ten
  mini-batches.}
  \label{tab:exploratory_pruning_protocol}
  \begin{tabular}{lccc}
    \toprule
    System & Ranking/evaluation source & Batch size & Maximum $q$ \\
    \midrule
    OMat24 perovskite, 1000\,K & same exported test loader & 10 & 60\% \\
    OMat24 perovskite, 3000\,K & same exported test loader & 10 & 60\% \\
    MPF2021 F--Fe--O & same 224-frame test loader & 10 & 60\% \\
    Aspirin (MD-17) & same molecular test loader & 20 & 60\% \\
    \bottomrule
  \end{tabular}
\end{table}

\end{document}